\documentclass[authoryear,final,5p,times,twocolumn]{elsarticle}
\usepackage{ifthen}
\newboolean{publ}
\setboolean{publ}{true}

\makeatletter
\def\ps@pprintTitle{%
  \let\@oddhead\@empty
  \let\@evenhead\@empty
  \def\@oddfoot{\footnotesize\itshape
       Printed in: \ifx\@journal\@empty Elsevier
       \else\@journal\fi
       \ 2011, 13(2):228--238; doi: 10.1111/j.1525-142x.2011.00472.x
       \hfill\today}%
  \let\@evenfoot\@oddfoot
}
\makeatother

\newcommand{\libpath}{/Users/Lussanet/Library/TeXShop/}

\usepackage{setspace} 
\usepackage{amssymb,amsmath,mathcomp,textcomp}	
\usepackage{graphicx}			
\usepackage{lineno}
\usepackage[colorlinks]{hyperref}	 
\usepackage{textcomp}			
\usepackage[applemac]{inputenc}

\begin{document}

\title{A hexamer origin of the echinoderms' five rays\tnoteref{t1}\tnoteref{t2}}

\tnotetext[t1]{This is the accepted, peer-reviewed version, without the Journal's editing. Some of the figures are color versions.} 
\tnotetext[t2]{Two corrections were made, see footnotes \ref{fnStalk} and \ref{fnFig5d}.}

\author[wwu]{Marc H.~E.~de Lussanet\corref{cor1}} 
\ead{lussanet@wwu.de}

\journal{Evolution and Development}

\address[wwu]{Westf. Wilhelms-Universit{\"a}t, Fliednerstra{\ss}e 21, 48149 M{\"u}nster, Germany} 

\ifthenelse{\boolean{publ}}{}{\onehalfspacing}

\begin{abstract}
Of the major deuterostome groups, the  echinoderms with their multiple forms and complex development are arguably the most mysterious. Although larval echinoderms are bilaterally symmetric, the adult body seems to abandon the larval body plan and to develop independently a new structure with different symmetries. The prevalent pentamer structure, the asymmetry of Lov{\'e}n's rule and the variable location of the periproct and madrepore present enormous difficulties in homologizing structures across the major clades, despite the excellent fossil record. This irregularity in body forms seems to place echinoderms outside the other deuterostomes. Here I propose that the predominant five-ray structure is derived from a hexamer structure that is grounded directly in the structure of the bilaterally symmetric larva. This hypothesis implies that the adult echinoderm body can be derived directly from the larval bilateral symmetry and thus firmly ranks even the adult echinoderms among the bilaterians. In order to test the hypothesis rigorously, a model is developed in which one ray is missing between rays IV-V (Lov{\'e}n's schema) or rays C-D (Carpenter's schema). The model is used to make predictions, which are tested and verified for the process of metamorphosis and for the morphology of recent and fossil forms. The theory provides fundamental insight into the M-plane and the Ubisch', Lov{\'e}n's and Carpenter's planes and generalizes them for all echinoderms. The theory also makes robust predictions about the evolution of the pentamer structure and its developmental basis.
\end{abstract}

\maketitle 
\ifthenelse{\boolean{publ}}{}{\linenumbers}

\section{Introduction}

The echinoderms display an exceptional variation in body forms. This is so for the extant clades, but even more so for the many extinct forms \citep{sumrall2007.149-163}. By virtue of their skeleton, the fossil record of echinoderms is well-known. Paradoxically though, this has not made the understanding of their evolution, or the relationships between the clades, easy \citep{hyman1955.book, david1998.21-28}. Moreover, the adult body form of echinoderms is apparently so disparate from other deuterostomes that it has been exceedingly difficult to develop a unified understanding of the deuterostome body. Although it has been clear from the embryological development for more than a century that echinoderms are deuterostomes (like vertebrates, cephalochordates, urochordates, and hemichordates), their pentaradial structure (as well as the lack of gills and chorda, and the stereom-structure of their skeleton) makes them the most atypical major lineage of deuterostomes. 

Important advances in the understanding of the echinoderm structure have come from detailed comparative paleontological studies \citep{david1996.577-584,mooi2005.542-555,hotchkiss1978.537-544,hotchkiss1998.200-214,sumrall2010.269-276}. Whereas the paleontological advances result from comparative studies of the extant and extinct echinoderm lineages,  molecular biology has focused mainly on a single model-organism, the sea urchin \emph{Strongylocentrotus purpuratus} \citep{mooi2008.43-62}. 

According to a theory by \cite{jefferies1986.book}, the calcichordate theory, the rays of echinoderms are derived from the pre-existing arms of a pterobranch-like ancestor (which putatively possessed more than five rays). Jefferies proposed that the reduction is a result of natural selection and the number of five rays is arbitrary.  However, in asteroids, in which the number of arms may vary markedly between species, it has been shown that a larger number than six rays is always secondarily derived from an underlying pentamery \citep{hotchkiss2000.340-354}.  Moreover, the calcichordate theory ignores the characteristic and unique stereom structure of the echinoderm ossicles, which are composed of calcite. Further, the current understanding of the skeletal structure of echinoderms (the extraxial-axial theory, see below) has shown that the presumed calcichordate fossils actually were true echinoderms and unrelated to chordates \citep{david2000.529-555}. Also, phylogenetic studies have shown that vertebrates and echinoderms are only distantly related within the deuterostomes \citep{halanych1995.72-76, delsuc2006.965-968,lartillot2008.1463-1472}.

Echinoderms have a complex development, in which the visible pentaradial structure appears late and is centered around the mouth. Many important, groundbreaking studies on the embryology were performed in the late 19th and early 20th century. These resulted in a great number of precise and vivid descriptions of the normal and abnormal embryology. This work has seen relatively little follow-up \citep{hyman1955.book, lacalli2000.421-432}. Echinoderm embryos develop directly or indirectly. In direct development, some developmental stages are very much shortened  \citep{Giese1991.book}. In its most complete indirect form, a free-swimming pelagic development is followed by attachment to initiate a sessile benthic stage, where the attachment may be released later, as in sea stars. In a stalked crinoid, the sea lily \emph{Metacrinus rotundus}, the blastula becomes an auricularia larva as in holothuroids, before developing into a doliolaria as in other crinoids and holothuroids \citep{nakano2003.158-160}. In the adult body, the left hydrocoel (mesocoel), left somatocoel (metacoel), and right somatocoel become arranged (stacked) on the oral-aboral axis \citep{peterson2000.93-101, david1998.21-28}.

A number of theories for the evolution of the pentaradial organization and body axes therefore see the adult axes as highly disparate from the larval axes \citep{hotchkiss1998.200-214, kerr1999.93-103, mccain1994.395-404, gudo2005.191-216, sumrall2007.149-163}. According to \citet{hotchkiss1998.200-214}, the pentaradial structure developed through the Bateson-type of duplication of the ray, from one to three. Subsequently, bilateral symmetry was re-established through the duplication of these secondary rays so that finally five rays are present. However, it is unlikely, or at least there is no consensus, that the uni-ray configuration is basal to echinoderms \citep{david2000.529-555,guensburg2001.131-134, sprinkle1998.269-282, sprinkle2004.266-280}. Instead, homalozoan forms have been interpreted as highly specialized adaptations, which are very old but nevertheless originate higher up in the echinoderm tree than pentaradial forms \citep{mooi1998.965-974,sumrall1997.267-288}. The PAR-model of \citet[Paedomorphic Ambulacra Reduction]{sumrall2007.149-163} proposed that the five-rayed structure is basal to the adult echinoderm \emph{Bauplan}, a symplesiomorphy. They proposed the three-ray as well as the uni-ray structure of the forms on which the calcichordate theory was based, are derived by reduction from five. Moreover, they propose that the pentaradial \emph{symmetry} in some recent clades is a derived feature. In favor of the rays-as-appendages model is the fact that further duplications of rays are common in the evolution of sea stars, whereas the early appearance of fossil pentaradial forms \citep{david2000.529-555} would be evidence for the PAR model.

\begin{figure*}[htb!]
\centering
\includegraphics[width=0.75\textwidth]{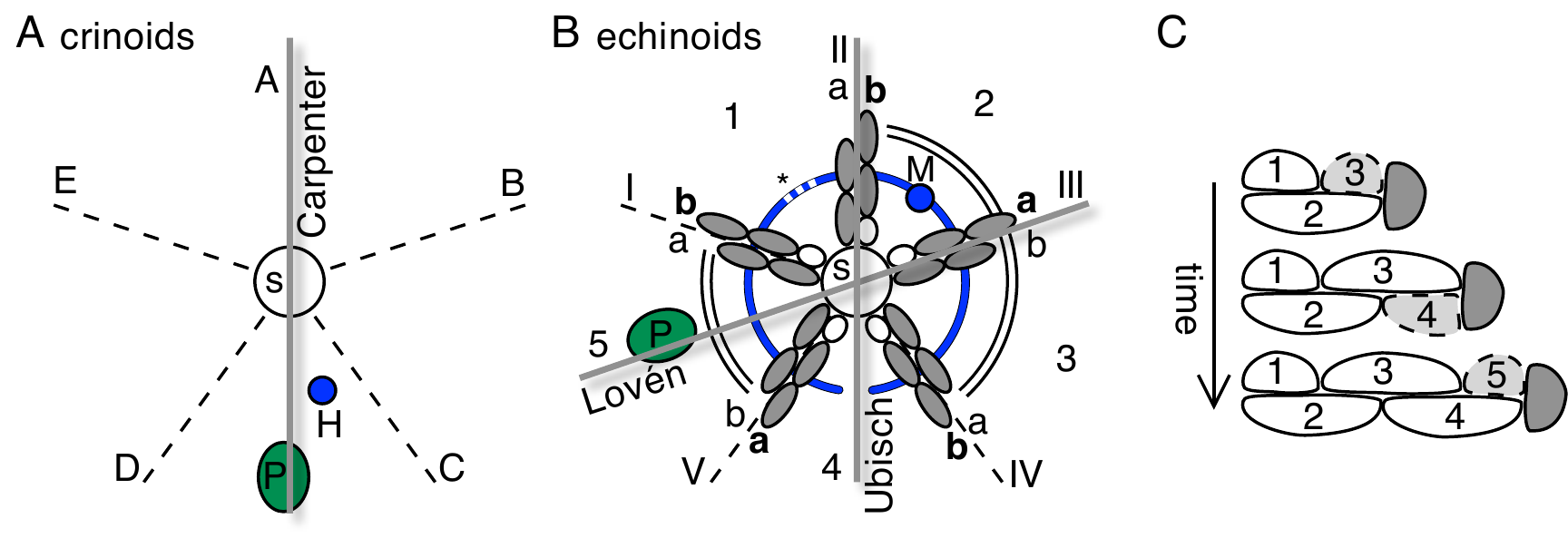} 
\caption{
(A) Carpenter's symmetry plane and encoding schema for the rays of crinoids (capitals A--E). s, mouth; P, periproct; H hydropore. (B) Lov{\'e}n's rule, discovered first for echinoids. Abbreviations of the rays: I--V refer to Lov{\'e}n's schema, whereas numerals 1--5 are the interradials \citep{loven1874.1-91, saucede2003.399-412}. The rows of ambulacral plates of a ray are referred clockwise by a and b respectively. The ontological basis for Lov{\'e}n's schema was developed by \citet{david1995.155-164}: the first tube-foot bearing ambulacral is respectively on the b, b, a, b, and a side (in short: bbaba). Lov{\'e}n's plane follows III--5, whereas von Ubisch' plane is aligned with  II--4 and crosses the closure point of the hydrocoel ring of the water vascular system (grey crescent). * in \emph{Holopneustes purpurescens} the early hydrocoel is divided in two pouches \citep[see text]{morris2007.1511-1516}. (C) The ocular plate rule for the formation of ambulacral floor plates in the axial region \citep{mooi1994.87-95}. New plates are added behind the distal ocular region, depicted in dark grey (this terminal blastema is mostly not ossified). The positions of the left and right ambulacrals are not always alternating, in asteroids and living ophiuroids they are opposite. In holothuroids, where the axial region is very much underdeveloped, no ambulacrals are formed in the actual rays --  the tentacles.} \label{fig1Intro}
\end{figure*}

\citet{kerr1999.93-103} explored the unusual flexibility of the \emph{Baupl\"ane} in holothuroids and found that bilateral symmetry must have evolved independently several times in holothuroids, and that this symmetry is independent of the bilateral symmetry of the larva and is of ectodermal origin. Internally, the pentaradial structure is robust. In one clade, the Rholpalodinidae, the mouth and anus fuse into a single tube and they develop an external decaradial symmetry, while retaining the internal pentaradial symmetry as well as the secondary bilateral symmetry. \citet{mccain1994.395-404} reversed the dorsoventral axis experimentally in four-cell staged echinoid embryos and found that this reversed the (embryonal) left--right axis as well, and thus the adult dorsoventral axis. \citet{gudo2005.191-216} proposed a model for the radial formation of rays as a result of hydrostatic pressure in the hydrocoel, a hypothesis that seems to be lacking empirical support. 

Nearly perfect pentaradial symmetry of the rays appeared late in the evolution of echinoderms, and only in the eleutherozoans [all extant echinoderms apart from crinoids] \citep{sumrall2007.149-163}. In other groups, adult echinoderms display at best an approximate bilateral symmetry and an approximate pentaradial symmetry, as will be discussed in more detail below. By contrast, echinoderm embryos  are typically bilaterally symmetric and  similar to hemichordate embryos, which has helped early embryologists to recognize echinoderms as deuterostomes. The mouth then moves (or disappears) to land (or re-appear) on the left side of the body. After that, axial structures associated with the hydropore form the circular coelom around the mouth with its five extensions. The coelom spaces are regrouped in a process called coelomic stacking \citep{david1998.21-28,mooi2005.542-555,peterson2000.93-101}. This is also illustrated below (Fig.\ \ref{fig5Model}D).  The pentamer structure seems to be imposed by the axial structures around the mouth on the rest of the adult body \citep{david1996.577-584}. Thus, in current thinking, the structure of the larva disappears almost completely in the transition to the adult body, which is assembled anew. 

However, even though a nearly perfect geometrical radial symmetry evolved late, the pentaradial structure was already established in early Cambrian times \citep{sprinkle1973.book}. Furthermore, although the organization is most prominent in the structures around the mouth, it is clearly present in the oral and aboral coelomic systems and nervous systems, and is as such deeply rooted in the echinoderm body \citep{david1996.577-584,mooi2008.43-62}. For example, of the three nervous systems even the aboral one is clearly and fundamentally pentaradial \citep{hyman1955.book,grasse1948.book}.

The present study presents a new hypothesis in which the pentamer structure is central to echinoderms. The hypothesis is fully consistent with the modern understanding of echinoderm skeletal structure as well as with the phylogenetic position of echinoderms. The hypothesis generates a number of predictions that should be testable in the framework of modern embryology and molecular biology. According to the hypothesis under consideration, the pentamer structure is a reduction from a hexamer structure. Since hexamery is common in bilaterians, the hypothesis provides a fundamental link between deuterostomes and the other bilaterians, and therefore offers new insights into the deuterostome and bilaterian body plan. Before outlining this hypothesis, I shall describe some fundamental features of the echinoderm structure.

\section{Symmetry and Lov{\'e}n}

\begin{figure*}[htb!] \centering
\includegraphics[width=0.8\textwidth]{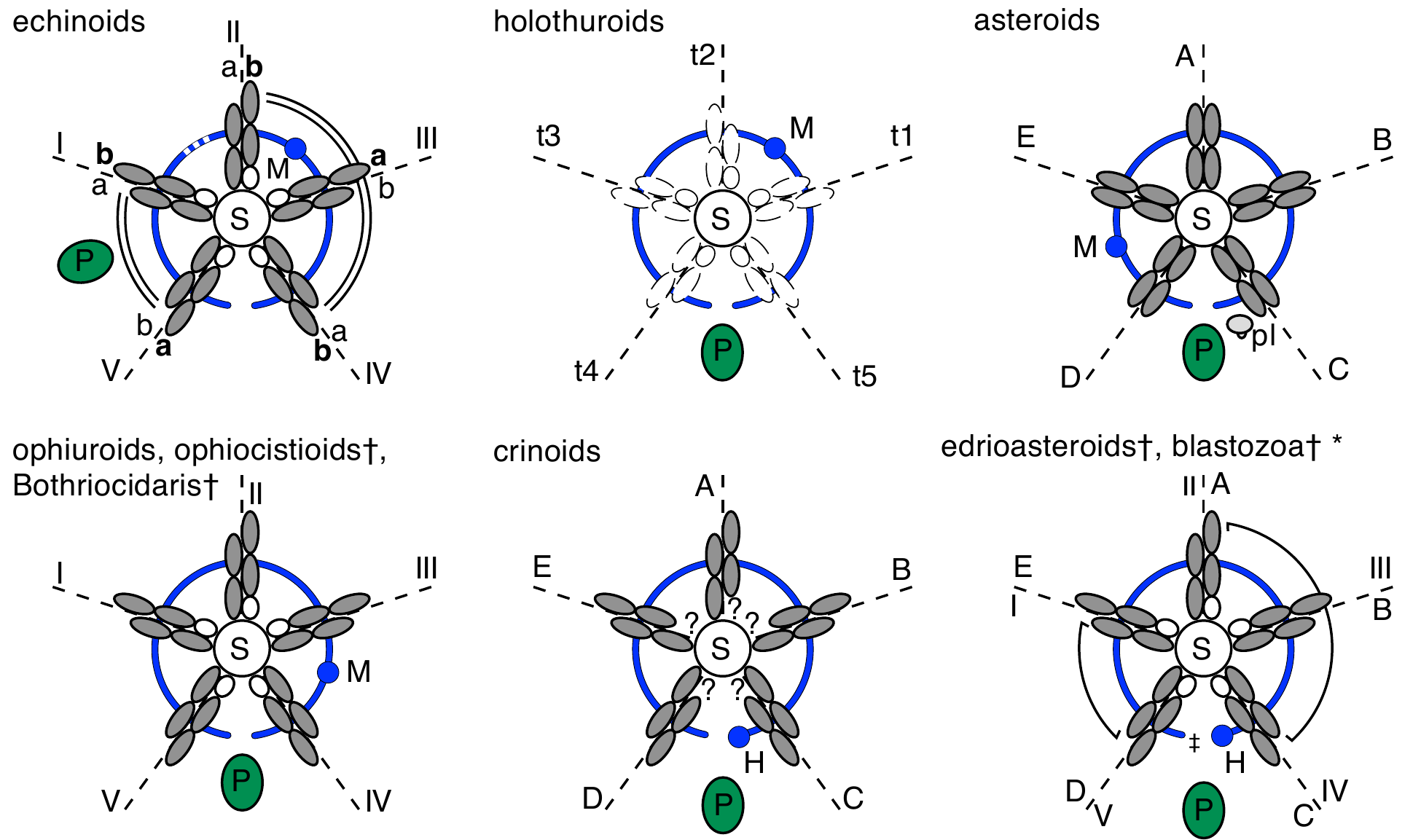}
\caption{
Schema of rays and body-openings in echinoderms, in oral (axial) view; based on Figure 2 of \citet[][and pers. comm. 2009]{hotchkiss1998.200-214}. Mouth S, madrepore M, or hydropore H, periproct P (which contains the anus).  Abbreviations of the rays: I--V Lov{\'e}n's schema, capitals A--E Carpenter's schema; the encoding of the rays follows  \citet{hotchkiss1998.200-214}. The blue crescent indicates the hydrocoel ring of the water vascular system. Figures are oriented with the closure of the hydrocoel ring downwards (von Ubisch's axis). 
\textbf{Echinoids}: Double arches: the posterior bivium and the anterior trivium in spatangoids, holasteroids, clypeasteroids, cassiduloids. Exocyclic P is only known for Irregularia. 
\textbf{Holothuroids}: location of P based on \citet{westheide2007.book}; despite the absence of ocular plates, there is some evidence for Lov{\'e}n's rule \citep{mooi2005.542-555}. 
\textbf{Asteroids} have symmetric rows of ambulacrals, but Lov{\'e}n's rule has been described \citep{morris2009.1277-1284}. In \emph{Luidia}, P is absent and the location of the M is deviant. Sucker/preoral lobe, pl. The \textbf{ophiuroids} lack P; in the \textbf{Bothriocidaris}\dag\ the location of M is deviant, P cannot be placed. 
For the \textbf{crinoids}, Lov{\'e}n's rule has not been proven (see text).  
In the \textbf{edrioasteroids}\dag\ and five-rayed \textbf{blastozoans}\dag\ the location of the closure of the hydrocoel ring (\ddag) is uncertain. The arcs indicate the grouping of the rays in \emph{Isorophusella incondita} \citep{bell1976.1-447}. Note, that the application of Lov{\'e}n's schema is deviant from the earlier literature for some edrioasteroids \citep[see][]{hotchkiss1998.200-214}. 
* Lov{\'e}n's rule is valid for five-rayed blastozoans, although the schema of the figure is not strictly correct \citep[glyptocystitoids$\dag$, eocrinoids$\dag$:][]{sumrall2008.229-242}.
} \label{fig2Structure} \end{figure*}

It has long been unclear whether any general rule or law can be formulated for the vast diversity of echinoderm skeletal plates \citep{hyman1955.book}. This is reflected in the large number of symmetry planes \citep{hotchkiss1995.401-435, gordon1929.289-334} proposed between and even within clades, and which is an ongoing source of confusion in the comparative literature. The embryological symmetry plane has already been mentioned. This \textbf{larval plane} is usually meaningless in the adult due to the extensive reorganization. Further, there is the \textbf{Carpenter plane} of crinoids that passes through  the mouth and the periproct (containing the anus). In \textbf{Carpenter's schema}, this is through ray A and interradial C-D (Fig.\ \ref{fig1Intro}A). The \textbf{M-plane} is defined for all echinoderms as the plane passing through madrepore (hydropore) and mouth \citep{bather1900.1-344}. This plane is highly variable with respect to the Carpenter plane (Fig.\ \ref{fig2Structure}). Unfortunately, there is in the literature a second ``Carpenter'' encoding schema which is defined with respect to the M-plane \citep{bather1900.1-344,hyman1955.book,morris2005.456-467}. In order to distinguish the schemas, I will refer to this encoding as the \textbf{Carpenter-Cu{\'e}not schema} \citep[acknowledging][]{cuenot1891.monografie}. Two planes have been found in echinoids (Fig.\ \ref{fig1Intro}B). The \textbf{Lov{\'e}n-plane} passes mid-way  through the posterior bivium and anterior trivium in asymmetric echinoids \citep{loven1874.1-91}. In \textbf{Lov{\'e}n's schema} rays are encoded with respect to this plane (Fig.\ \ref{fig1Intro}B). Symmetric echinoids can be fitted into this schema on the basis of Lov{\'e}n's rule (see below). \citet{ubisch1913.119-156} derived a primordial axis on the basis of echinoid embryology. This \textbf{Ubisch plane} passes through the mouth and the closure of the hydrocoel crescent \citep{hotchkiss1998.200-214}, and aligns Lov{\'e}n's rule for edrioasteroids, ophiuroids, holothuroids, and echinoids  \citep{david1995.155-164}.

The homologies of the skeleton of echinoderms are now well understood, and do follow a few, general, rules \citep{mooi2005.542-555, sumrall2010.269-276}. According to the extraxial-axial theory (EAT), these rules have an ancient genetic-ontological basis \citep{mooi1994.87-95, mooi2005.542-555, mooi2008.43-62} and are valid even in long-extinct clades \citep{david2000.529-555}. According to the EAT, the echinoderm body is divided into the \emph{axial region} containing the mouth, the \emph{extraxial region} which contains the other body openings (perforate-extraxial), and the \emph{imperforate-extraxial region} or stalk-region which is present in only some clades. A stalk has probably evolved several times. The formation of skeletal plates in the extraxial region of echinoderms follows no generally valid rule, but the development of the axial region in all echinoderms follows two general rules. 

In the axial region, the pentaradial skeletal structure is most prominent. The ambulacral skeletal plates in this region grow in five directions away from the mouth, following the \emph{ocular plate rule} (OPR). New plates are added in two rows adoral to the terminal ocular region, alternating on the left and on the right side (Fig.\ \ref{fig1Intro}C). Usually, the first ambulacral plate of each ray is smaller than its successors so that the ambulacrals often form an alternating pattern. In two rays the first ambulacral plate is in the left and in three rays it is in the right row of plates. The pattern of left-starting and right-starting rays was discovered by Lov{\'e}n for echinoids \citep{loven1874.1-91,ubisch1913.119-156,david1995.155-164}. According to this rule (\textbf{Lov{\'e}n's rule}), the arms are ordered with the first plate right, right, left, right, and left side of the biserial arrangement, counting from ray I \citep[or: bbaba,][see Fig.\ \ref{fig1Intro}B and \ref{fig2Structure}]{hotchkiss1978.537-544, hotchkiss1998.200-214, david1995.155-164}. It has long been unclear whether Lov{\'e}n's rule also applies to other clades, but now it has been described in ophiuroids, ophiocistioids, and in the fossil edrioasteroids and five-rayed blastozoans \citep{hotchkiss1995.401-435, sumrall2008.229-242}. Lov{\'e}n's rule has recently been described for the development of the primary podia of the sea star \emph{Parvulastra exigua} \citep{morris2009.1277-1284}. In crinoids, Lov{\'e}n's rule has not been proven. Since it is still debated whether crinoids stem from edrioasteroids or blastozoans \citep{guensburg2009.350-364}, the interpretation of the ambulacral floor plates cannot be decided yet. Nevertheless, Lov{\'e}n's rule is now accepted to be the ancestral echinoderm condition. 

As a result of the EAT, an underlying structure and order in the diversity of echinoderm body forms is now recognized, which incorporates even the most atypical forms, the polyphyletic ``homalozoans'' which possessed just a single arm instead of five \citep{david2000.529-555,lefebvre2003.511-555}. 

In crinoids as well as edrioasteroids, a basal group of echinoderms, rays (D+E)--A--(B+C) form a 2--1--2 arrangement that is almost bilaterally symmetric with respect to the mouth and periproct (Fig.\ \ref{fig2Structure}). There are several extinct forms in which the mouth is not located in interradial C--D, e.g.\ in paracrinoids \citep{sumrall2009.135-139} and glyptocystitids \citep{sumrall2002.918-920} the periproct is located in interradial B--C. In echinoids, this 2--1--2 arrangement is concealed by the 2--3 or bivium-trivium arrangement of rays (V+I)--(II+III+IV) in spatangoids, clypeasteroids, holasteroids, cassiduloids. Likewise there is a 2--3 grouping of rays (D+E)--(A+B+C) in the Edrioasteroid \emph{Isorophusella incondita} (Fig.\ \ref{fig2Structure}). The rays in these 2--3 groupings are homologous based on the expression of Lov{\'e}n's rule in \emph{Stromatocystites} and in echinoids \citep{hotchkiss1998.200-214}. A consequence of this homology is that Hotchkiss' application of Roman numerals to edrioasteroids, as replicated in Figure \ref{fig2Structure}, is different from prior practice (the anterior ray was formerly labelled III). A bivium-trivium pattern also appears in the early development of the hydrocoel in the echinoid \emph{Holopneustes purpurescens} \citep{morris2007.1511-1516}. She found that the early hydrocoel is divided into two pouches which develop respectively two and three primary podia. Morris applied the Carpenter-Cu{\'e}not encoding to the rays, but using the Lov{\'e}n schema instead, it becomes clear that the arrangement of the hydrocoel pouches conforms with the bivium--trivium grouping of rays.

\begin{figure}[htb!] \centering 
\includegraphics[width=\columnwidth]{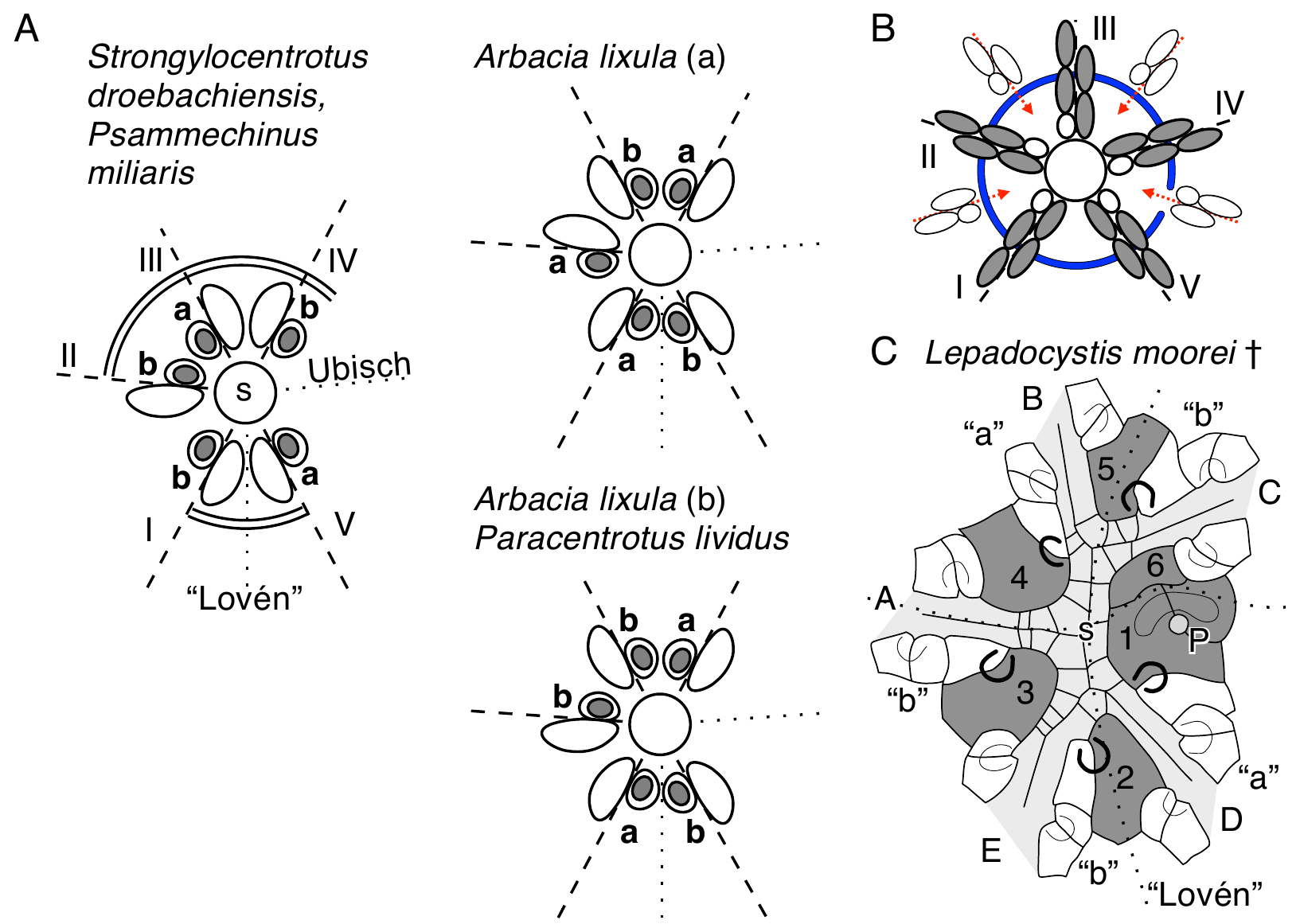}
\caption{(A) The findings of \citet{ubisch1913.119-156} and \citet{gordon1926.259-312} for the primary tube feet on the buccal plates of the larvae of four echinoid species. Grey circles indicate the position of the primary tube feet. Von Ubisch stressed that ray II breaks the symmetry. Here, the figure is drawn so as to reveal the symmetry of the other rays with respect to ``Lov{\'e}n's plane'' (quotes to acknowledge that the plane is usually drawn through ray III). Note that von Ubisch referred to \emph{S.\ droebachiensis} as \emph{Toxopneustes dr\"obachiensis}, and to \emph{P. lividus} as \emph{Strongylocentrotus l.}; his \emph{A. pustulosa} is probably \emph{A. lixula}; Gordon referred to \emph{P. miliaris} as \emph{Echinus m.} 
(B) The four possibilities to restore symmetry into Lov{\'e}n's rule by adding a ray. 
(C) \citet{sumrall2008.229-242} showed that Lov{\'e}n's rule can be applied to the primary brachioles of pentaradial blastozoid echinoderms ($\dag$). Brachiole facets are depicted as arcs. Here, a glyptocystitoid is drawn in the orientation of panel A. Axial plates are drawn light grey, biserial floor plates plates white. Dark grey are oral plates (numerals 1--6). The quoted ``a'', ``b'' refer to the position of the primary brachioles (fat arcs). After \citep{sumrall2008.229-242}.  Abbreviations: see Fig.\ \ref{fig1Intro}B.} 
\label{fig3Ubisch} 
\end{figure}

The echinoid mouth is surrounded by buccal plates, which may be located proximally to the ambulacral plates. In regular acroechinoids, the first ten buccal plates  appear in pairs, one of which carries a central tube foot, whereas the other acquires one later \citep{gordon1926.259-312}.  \citet{david1995.155-164} showed that the buccal plates bearing a tube foot right from first appearance are in fact modified ambulacral elements. \citet{loven1874.1-91} found for larval \emph{Strongylocentrotus droebachiensis} (syn. \emph{Toxopneustes dr\"obachiensis}) that his rule also applies to the tube feet, in the sense that the appearance of the first primary tube foot in each ray is the same as that of the first ambulacral plates (Fig.\ \ref{fig3Ubisch}A). \citet{ubisch1913.119-156} confirmed this finding, as well as did \citep{gordon1926.259-312} for \emph{Psammechinus microtuberculatus} (syn.\ \emph{Echinus m.}) and \citet{morris2009.597-608} for \emph{H. purpurescens}, but failed to reproduce Lov{\'e}n's rule for the tube feet of two other species. Instead, von Ubisch found that the rule was inverted in one \emph{Arbacia} larva (Fig. \ref{fig3Ubisch}A). Even more confusingly, the larvae of \emph{Paracentrotus}, and another \emph{Arbacia} displayed an inverted pattern for ray II. (Von Ubisch also observed a \emph{P. microtuberculatus} pluteus larva with two tube feet on both buccal plates of ray II, but this larva was probably older than the one for which \citet{gordon1926.259-312} did report the bbaba arrangement of the primary tube feet of this species.) 

\section{A Hexamer Structure}

On the basis of the above finding, von Ubisch rejected Lov{\'e}n's plane in favor of his primordial plane. However, although Lov{\'e}n's rule is \emph{more} symmetric with respect to the Ubisch plane than with respect to Lov{\'e}n's plane, it is still asymmetric. One way to restore symmetry into Lov{\'e}n's rule is to assume that one ray is missing. Figure \ref{fig3Ubisch}B shows that symmetry can be restored in four different ways by adding one ray. This shows, that the asymmetry of Lov{\'e}n's rule might be derived from a bilaterally symmetric hexamer rule. Adding 
one ray 
between rays I-II or IV-V would restore Lov{\'e}n's symmetry plane into b\textbf{a}b$|$aba and bba$|$b\textbf{a}a respectively. On the other hand, adding one ray between I-II, II-III or III-IV would result in bilateral symmetry with respect to von Ubisch' primordial plane (b\textbf{a}$|$bab$|$a, bb$|$\textbf{a}ab$|$a, or bb$|$a\textbf{a}b$|$a respectively).

As said above, von Ubisch found that ray II is the one that is not paired in Lov{\'e}n's rule. Ray II can be paired in two ways: by adding a ray between I-II or IV-V, both of which would provide bilateral symmetry approximately with respect to Lov{\'e}n's plane. This result is paradoxical, since von Ubisch introduced his approximate symmetry plane to overcome the asymmetry of Lov{\'e}n's rule. 

Which of these two solutions is supported by data? A possible answer is provided by the model of \citet{sumrall2008.229-242,sumrall2010.269-276} for blastozoan echinoderms, as illustrated for \emph{Lepadocystis} (Fig.\ \ref{fig3Ubisch}C). As explained above, Lov{\'e}n's rule describes the position, in biserial row a or b, of the first ambulacral plate that bears a tube foot \citep{david1995.155-164}. In blastozoans the ambulacra were flanked by food-gathering appendages, called brachioles. These brachioles were located on the suture between two floor plates (arches in Fig.\ \ref{fig3Ubisch}C).  One of the plates bearing the proximal brachioles is always an oral plate (bold arches, dark grey plates). The location of these primary brachioles follows Lov{\'e}n's rule \citep{sumrall2008.229-242}. 

According to the model of Sumrall, the peristomial area of blastozoans is flanked by six oral plates, because one extra plate is present in interradial C-D. To illustrate this more clearly, \emph{Lepadocystis} is aligned according to Lov{\'e}n's plane (Fig.\ \ref{fig3Ubisch}C), whereas some extra space is left free between the IV-V rays of the echinoderms in panel \ref{fig3Ubisch}A. The relation between the primary tube feet and the primary brachioles is not known. Since the oral plates, which are extraxial elements, bear brachioles, the latter should probably be regarded as extraxial elements too. If the application of Lov{\'e}n's rule to blastozoans is correct then this would mean that the probable location of a missing ray is between rays C and D, the only interradial which  bears two oral floor plates that flank the oral region. 

This result has four important consequences. First, it suggests that the axial elements (of the EAT) are derived from a bilateral symmetry, whose plane coincides approximately with Lov{\'e}n's, that is, through the mouth and interradials 3 and 5. Second, it follows that the element missing from this symmetry is located opposite to ray II. Third, this bilateral symmetry is consistent with the bivium-trivium compartmentalization, by supplementing it into a bivium-quartium. Fourth, the symmetry is inconsistent with an underlying triradiate structure as proposed by some of the theories mentioned in the introduction.

\section{An Evolutionary and Developmental Model}

Two crucial questions arise: Why should one ray have been lost in the early evolution of echinoderms, and how does this occur developmentally? We start with the first question by defining a hypothesis, namely that the sixth ray has been reserved for settlement, in the transition from a pelagic larva to a sessile, benthic stage. Such a transition still occurs in two of the major extant lineages, viz.\ the Crinoidea and Asteroidea, and probably in all the extinct sessile clades. In asteroids, the sessile stage is usually very brief, but all species seem to produce a 
sucker or preoral lobe.%
\footnote{The original erroneously said ``stalk''.\label{fnStalk}}

The answer to the second question is based on this settlement hypothesis and aims to answer the following important questions. How can the pentamer adult body be derived from the bilateral symmetry of the larva? How does the position of the stalk fit in? Why does the stalk of crinoids, edrioasteroids and blastozoans not contain axial elements if the missing element is caused by producing a stalk? We will see that the model can indeed answer these questions and even suggest explanations for some extinct uni-rayed forms and the variant locations of the hydropore and periproct with respect to Lov{\'e}n's rule.

Central to the model that we will develop is the following scenario. Imagine a pelagic larva that will settle with its mouth turned downward. If it is to feed like a sessile echinoderm, the mouth should move up, away from the part of the body where it settles. Crinoids settle on the adhesive pit, and asteroids settle on the adhesive disk (from which the sucker or preoral lobe develops, see Fig.\ \ref{fig5Model}E), both of which initially are located on very similar positions on the antero-ventral side of the larva \citep[][p. 132]{lacalli1993.127-133}. Subsequently, the mouth (or structures of the mouth region) moves to the left side of the body, where it meets the structures surrounding the hydrocoel. In the view of the model, it is the interaction between the ventral and the left-lateral structures of the hydrocoel from which the axial structures develop. 

It must be stressed here that this interaction is a crucial step in the development. It is very important to realize that the axial pentamer structure cannot originate from the hydrocoel on the left side of the body because it would have to develop de-novo. Instead, this structure must have been inherited from markers that moved with the mouth (or mouth-related structures if the mouth closes) to the left side of the body, because this already had a six-minus-one structure centered around the ventral mouth. This is most clearly reflected by the six coelom spaces. 

We have thus arrived at a model where one segment of the larva parts from the hexamer structure for settlement. Subsequently, ventral structures  migrate to the left side of the body where they imprint the (now pentamer) structure on the hydrocoel-related structures. What is still missing is the relation to Lov{\'e}n's rule. Now, an issue arrises that is truly puzzling at first sight. Above we saw (cf.\ Fig.\ \ref{fig3Ubisch}A), that the approximate Lov{\'e}n-plane (passing through interradials 3-5) forms the bilateral symmetry plane of the axial structures in the adult animal. However, I demonstrated above (cf.\ Fig.\ \ref{fig3Ubisch}A,C) that one ray is missing on the \emph{right} side, which is the direction in which the mouth has migrated. This appears to contradict the thesis that the mouth should move away from the side of settlement.

\begin{figure}[htb!] \centering
\includegraphics[width=0.7\columnwidth]{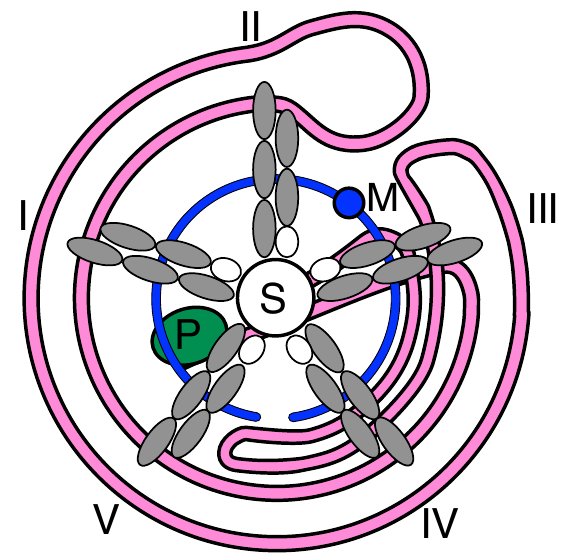}
\caption{The intestinal undulations in the echinoid \emph{Tripneustes ventricosus} \citep[after][]{hyman1955.book}.}
\label{fig4Intest}  \end{figure}

\begin{figure*}[htb!] \centering
\includegraphics[width=.9\textwidth]{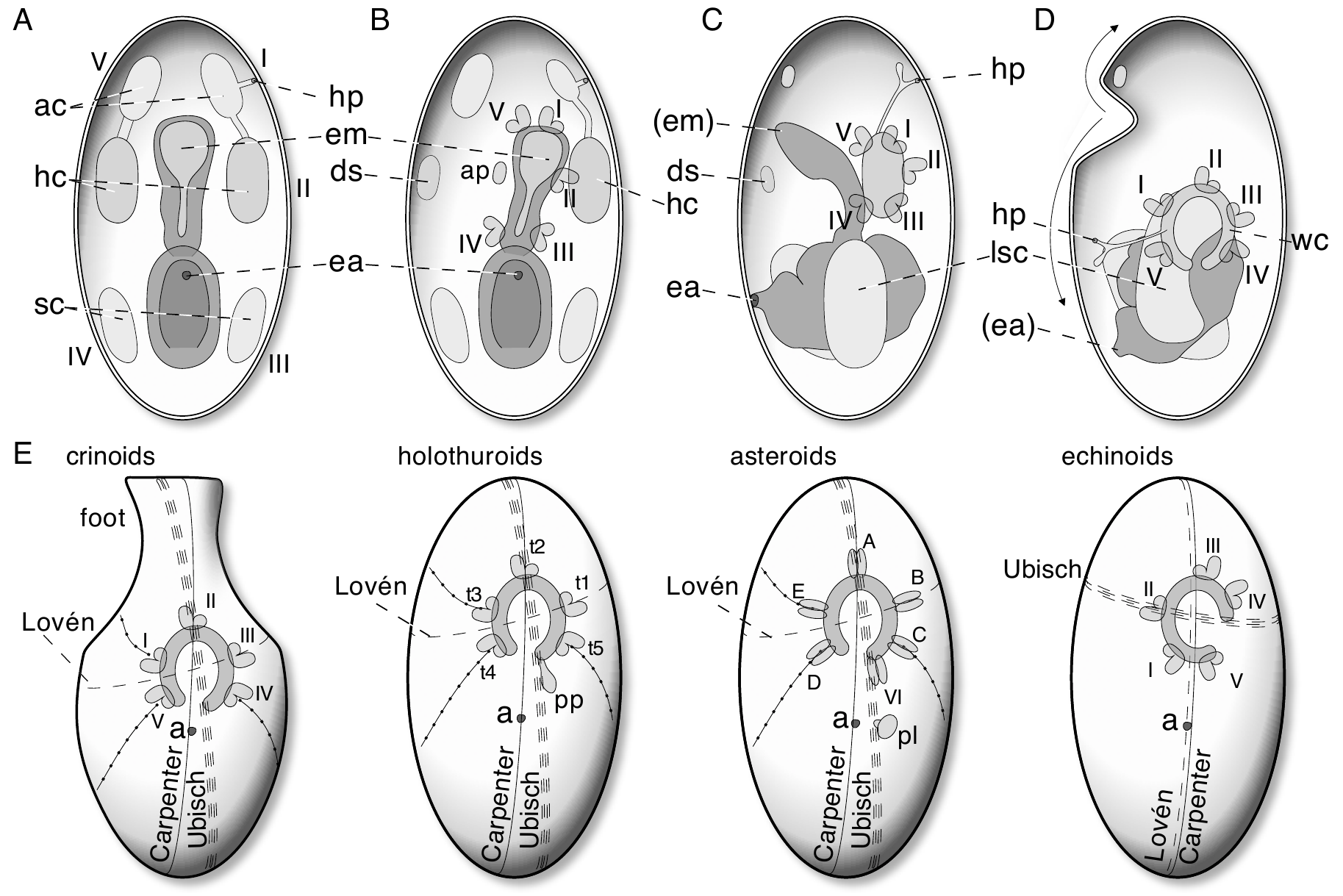}
\caption{General model of echinoderm development.
(A) Ventral view into the bilaterally symmetric stage of an echinoderm embryo, with six coelom spaces and a digestive tract. Abbreviations: ac, hc, sc: axo-, hydro, somatocoels; ea, em: embryonic anus, mouth; hp: hydropore. Roman numerals depict the proposed coelomic origin of Lov{\'e}n's schema. 
(B) Formation of the dorsal sac (ds) from the right hydrocoel and the opposite lateral movement of ventral structures surrounding the mouth. Supposedly, the mouth region is imprinted by the underlying coelomic structure (indicated as curved shapes), except for the right hydrocoel
. 
(C) Schematic lateral view depicting how the presumed ventral structures drift over the left hydrocoel. Formation of adhesive pit (ap). 
(D)  Formation of the water canal (wc), closing towards the missing symmetry. It moves towards the left somatocoel (lsc), coelomic stacking. Opposite rotation of the anterior and the central-posterior structures (arrows). 
(E) Schematic oral views of some extant clades, with body planes and anus (a) indicated. \textbf{Crinoids} develop the foot (and stalk) at the location of the ap (anterior in the larva). \textbf{Holothuroids} develop the primary polian vesicle (pp) at the location of the missing ray, whereas six-rayed \textbf{asteroids} and ophiuroids develop their sixth ray (VI) at this location. Notice the location of the sucker or preoral lobe (pl) which develops from the adhesive disk (panel C), see text. In \textbf{echinoids}, the axial structure rotates even further, inverting Lov{\'e}n's axis with respect to the larva.} 
\label{fig5Model}  \end{figure*}

The explanation for this apparent contradiction is be found in the trajectory of the intestinal tract inside the body, which revolves around the oral-aboral body axis. This is illustrated for an echinoid (Fig.\ \ref{fig4Intest}) where the twisting is especially pronounced. I shall analyze this particular schema below. 

In the echinoderm larva, the clearest sign of a hexamer organization is in the coelom spaces (Fig.\ \ref{fig5Model}A). One of these spaces does indeed disappear in eleutherozoans \citep{smiley1986.611-631}: the right mesocoel/hydrocoel, or dorsal sac, moves dorsally  before it disappears (Fig.\ \ref{fig5Model}B). In crinoid embryos, the right mesocoel and axocoel do not develop \citep{holland1991.247-300,nakano2003.158-160}. In the ophiuroid \emph{Ophiocoma nigra}, the right hydrocoel hardly develops and soon closes  \citep{narasimhamurti1933.63-88}. Asteroids, as observed by \citet{gemmill1914.213-294} in \emph{Asterias rubens}, sometimes develop two hydropores, that is, an additional one on the right side of the larva. Since these embryos nevertheless form a dorsal sac, it is not derived from the right axocoel, but from the right hydrocoel. In echinoids and asteroids \citep{macbride1903.285-330,bury1895.45-135}, the dorsal sac moves dorsally (away from the mouth) and remains throughout life. 

In a left-lateral view it is depicted how the ventral imprinting supposedly overlays the left hydrocoel region  (Fig.\ \ref{fig5Model}C). Subsequently, coelomic stacking occurs and the hydrocoel forms a crescent, developing five primary podia (Fig.\ \ref{fig5Model}D).%
\footnote{Unfortunately, panel \ref{fig5Model}D contained some errors and had to be replaced: in the original, the location of the opening of the hydrocoel crescent was mis-placed with respect to the encoded ray structure.\label{fnFig5d}}%
As the figure depicts, the hydrocoel rotates away from the anterior body-side. This rotation is most clear in crinoid embryos \citep{mladenov1983.309-323}. A very similar twisting has been described for the asteroid \emph{Solaster endeca} \citep{gemmill1912.book}. 

\begin{figure*}[hbt!] \centering
\includegraphics[width=.9\textwidth]{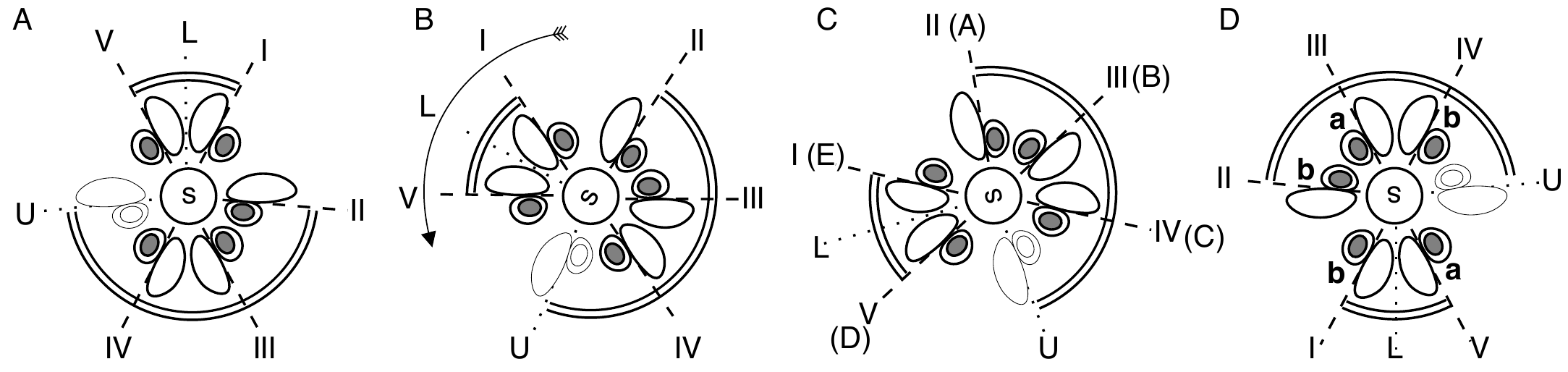}
\caption{The hypothetical rotation of the axial structure that leads to the orientation of the planes of Lov{\'e}n (L) and von Ubisch (U) in the adult. (A) bilaterally symmetric larva (ventral view, with anterior upward). (B) Left view, after migration towards the left side of the body. The direction of rotation is indicated with an arrow. (C) Adult echinoderm (apart from echinoids). (D) Adult echinoid. I-V: Loven's encoding; A-D: Carpenter's encoding.} 
\label{fig6Twist}  \end{figure*}

Notice that because of this twisting, the ventral structures have rotated around the antero-posterior axis as well as around the bilateral axis. The first twist is well-known and is the reason why the adult oral side is located on the larva's left. The second twist, according to the model developed here, is the reason for the confusion of symmetry planes in echinoderms. This is explained in Figure \ref{fig6Twist}, which schematizes an axial (oral) view while disregarding the migration towards the left side of the body. In panel A, the hypothetical orientation in the larva is shown in ventral view with the anterior pole upwards. The right ``ray'' never develops. According to this model Lov{\'e}n's plane is derived from the larva's sagittal plane, whereas Ubisch' plane is derived from the larva's transverse plane. Once the axial structure has reached the left side of the body, it rotates leftwards (anti-clockwise) around the axis perpendicular to the body surface (\ref{fig6Twist}B), until von Ubisch' plane is oriented approximately along the larval anterior-posterior axis (Fig.\ \ref{fig6Twist}C, cf.\ Figure \ref{fig1Intro}B, \ref{fig5Model}E). 

Figure \ref{fig5Model}E illustrates some variations on the general pattern. As shown by embryological development, the crinoid foot/stalk develops on the larval anterior pole \citep{mladenov1983.309-323, nakano2003.158-160,lacalli1993.127-133}. After the process of coelomic stacking, three rings of coelomic origin are aligned on the oral-aboral axis (cf.\ Fig.\ \ref{fig5Model}D), develop into circular coeloms, and each forms extensions into the five rays. The hydrocoel canal often forms one or more additional extensions in the interradials, such as the polian vesicles, best known in holothuroids \citep{nichols1966.219-244, nichols1972.519-538}. The primary polian vesicle is always formed on the t5-side of the hydrocoel crescent (Fig.\ \ref{fig5Model}E), \citep{bury1895.45-135, bury1889.409-450, selenka1883.monographie}.

Further evidence that the missing ray is indeed on the Ubisch plane comes from asteroids. Asteroids, like crinoids, form an adhesive disk in a very similar anterior-ventral location (see above). However, the sucker with which asteroids attach is formed between rays C-D (Fig.\ \ref{fig5Model}E). A simple explanation for this is that sea stars have their mouth not upward but downward, so the adhesive disk and mouth need not move to opposite directions to the body. \citet{gemmill1912.book} gives a vivid description of the complex twisting movements by which the anterior lobe of the sea star \emph{Solaster endeca} folds towards the left side while twisting clockwise around its axis (when looking at the anterior side). This motion is consistent with Figure \ref{fig5Model}C-D. However, in the meantime the preoral lobe where the sucker develops keeps its position with respect to the rays. This means that the preoral lobe does not move away from the mouth but indeed shifts together with the axial structure towards the posterior side of the body. 

If indeed the preoral lobe of sea stars never really leaves its location in the hexamer structure, then it may even form a true ray. Accordingly, the model predicts that if a sea star develops a sixth ray, then it should be located in the C-D interradial. Many sea stars can develop six rays or more. \citet{hotchkiss2000.340-354} analyzed the order in which the rays appear. In all cases where reliable data were available, a five-rayed stage was recognizable. Moreover, in all these cases the subsequent arms were formed after a pause, so Hotchkiss could show that the multi-rayed forms are derived from a pentamer Bauplan. Relevant for the prediction of the model is the location of the sixth arm. Indeed, the sixth arm appears in the C-D interradial in almost all cases. Moreover it appears on the dorsal (C) end of the hydrocoel crescent, and appears usually markedly earlier than its successors (if any) \citep{hotchkiss2000.340-354}. 

This was found for the six-rayed \emph{Leptasterias}, as well as for \emph{Pycnopodia}, \emph{Stichaster}, \emph{Crossaster}, \emph{Solaster}, \emph{Patiriella}, \emph{Acanthaster}, and \emph{Heliaster}.  In \emph{Cross\-aster papposus}, which typically has 12 arms, the hydrocoel canal of the sixth arm develops even prior to that of ray D. The canals of subsequent rays do not appear until two days later, on day 17 \citep{gemmill1920.155-187}. \citet{ritter1900.247-274} found in the 20-rayed \emph{Pycnopodia helianthoides} that the sixth ray develops at the location of the preoral lobe, and that the sixth ray develops markedly earlier than the following rays. An exception is formed by multi-rayed \emph{Luidia} species \citep{komatsu1994.327-333,horstadius1926.8,hotchkiss2000.340-354}. In these species, the supernumerary rays do not appear in the C-D interradial. For \emph{Luidia ciliaris}, H\"orstadius suggested that the aberrant development might be related to the different development of the hydrocoel which does not form a crescent but which is perforated by the mouth very late in development.

Finally, as illustrated in Figure \ref{fig6Twist}D, the rotation of the hydrocoel crescent is even more pronounced in echinoids than in any other echinoderm. As a result, Lov{\'e}n's plane coincides with Carpenter's plane in this clade. Paradoxically though, the plane is reversed by 180\textdegree\ with respect to the anterior-posterior axis of the larva. This is a prediction from the model, but it is consistent with the shape of the intestinal tract of echinoids (Fig.\ \ref{fig4Intest}). The intestinal tract makes two recoil loops around the oral-aboral axis. According to the model, the axial structure is rotated counter clockwise by a half revolution. The inner recoil loop crosses rays III, IV, and the Ubisch' plane that marks the missing ray, and thus confirms the model prediction. 

\subsubsection*{Hydropore / Madrepore}

One of the most confusing facts of echinoderms is the wandering position of the hydropore/madrepore (Fig.\ \ref{fig2Structure}). Especially surprising in this respect is that the hydropore directly communicates with the hydrocoel and that this connection exists already long before the mouth migrates leftward. To my knowledge there currently is no theory that even tries to explain this. Although the present model was not developed to explain the hydropore position, it readily provides a simple explanation that matches the data well. 

The second recoil loop of the intestinal tract of echinoids (Fig.\ \ref{fig4Intest}) is located more aborally than the first and makes a full revolution up to the interradial containing the madrepore. If we follow the same logic, this suggests that the aboral structures have rotated clockwise by a half revolution. This may seem surprising, but it is also consistent with the model, according to which the anterior region (containing the axocoel and the hydropore) rotates clockwise, opposite to the anti-clockwise rotation of the hydrocoel region with Loven's schema.


\section{Discussion}

The hexamery hypothesis thus might explain the evolution of echinoderm pentamery. The hypothesis unites the seemingly irregular echinoderm forms and provides them with a true bilateral symmetry axis. On the basis of this hypothesis, and to test it rigorously, a general model was developed to explain how this symmetry is lost in the adult animal as a result of dramatic twisting movements at the time of the pelagic-benthic transition. 

The hexamer hypothesis and the model are fully consistent with the extra\-xial-axial theory \citep{mooi2008.43-62} and the blastozoan model \citep{sumrall2008.229-242,sumrall2010.269-276}. In line with these models, the hexamer hypothesis requires that the pentamer structure is basal to echinoderms. Moreover, it provides a plausible mechanism by which this might have occurred. Since the aberrant uni-rayed echinoderms, the polyphyletic homalozoans, have been explained within the EAT as reduced from a pentamer structure \citep{david2000.529-555}, these clades do not present a problem to the hexamery hypothesis any more than to the EAT.

The present study exposes a paradox of Lov{\'e}n's plane that was suggested also by \citet{saucede2003.399-412}. Whereas Lov{\'e}n's rule is a feature of the axial structures (including the hydrocoel crescent), Lov{\'e}n's plane applies to the perforate extraxial body, because in echinoderms the plane intersects with the anus (periproct). The primary plane \emph{seems} to be von Ubisch' axis \citep{saucede2003.399-412}. However, as Sauc\`ede et al.\ stressed, this does not invalidate Lov{\'e}n's plane. Instead, the present work suggests that Lov{\'e}n's plane really is derived from the anterior-posterior axis of the larva (note that it is perpendicular to the sagittal plane of the larva because the mouth develops on the left of the larva). The reason that in echinoids Lov{\'e}n's plane intersects with the anus is that the plane is turned more strongly than in other echinoderms, by a half revolution.

The model predicts that the same genetic organizing mechanisms should be activated before and after the coelomic stacking, which indeed seems to be the case \citep{mooi2005.542-555,mooi2008.43-62}. Moreover, it predicts that \emph{Hox} gene expression should be present in the stalk region of crinoids. Indeed \citet{hara2006.797-809} found that \emph{MrHox5} is expressed in the anteriormost region of the somatocoels and initiates the stalk-related structures.

It is important to stress that the model does not necessarily predict that the stalk of sessile echinoderms, such as edrioasteroids, blastoids, and crinoids should display axial features. On the contrary, it is generally believed that the axial structures are related to the hydrocoel, which is located on the left side of the larva. The model proposes that the pentamery originates from the ventral side of the body. The axial structure might thus result from the interaction between ventral and left-lateral markers. 

We have thus reached the goal of the present work, in ranking the structure of the echinoderms firmly inside the bilaterians by providing them with a true bilateral structure. This structure is surprisingly close to the derived pentaradial one. The hypothesis presented here proposes that the pentaradial structure is an adaptation to the transition of a pelagic larval stage to a sessile stage. The hypothesis implies that the underlying structure of the adult echinoderms is derived from bilateral symmetry and shows a hexamer structure. This may seem counterintuitive at first sight. For example, a hexamer structure is not generally recognized to underlie the deuterostome bauplan. However, a hexamer structure superimposed on the bilateral symmetry is not unusual for bilaterians (e.g.\ annelids, nematodes). Also, the deuterostome larva at first develops six coelom spaces, the paired proto-, meso- and metacoels. In the coelomic stacking in echinoderms, this hexamer structure is lost and three of the coelom spaces encircle the intestinal tract. The hexamer structure is established after coelomic stacking, and this may be driven by simple molecular mechanisms. If both the mouth aperture and molecular signaling systems migrate from the ventral to the left side of the body, the hexamer structure can re-establish itself through the axial structures. This would explain how the pentamer (or in the light of the presented hypothesis: six-minus-one) is initiated at the mouth. 

The model proposes that a precisely defined and precisely timed, complex sequence of rotations occurs during larval development. Such movements were almost impossible to measure with the techniques available at the time of the classical developmental studies, but should be allowed by modern three-dimensional tracing techniques \citep{keller2010.637-642}. Obviously, the six coeloms themselves do not move to the left side of the larval body. Rather, it seems that cellular structures share the six-fold structure with the original six coeloms, and this cellular structure migrates to the left side. These important issues will have to be resolved in further research.


\section*{Acknowledgements}

I am indebted to Rich Mooi, Fred H.C.\ Hotchkiss, Bruno David, and an anonymous reviewer for their discussions, explanations, and valuable critique. I thank Kim Bostr{\"o}m, Guy Loosemore, and other colleagues for correcting the language.


\section*{References}

\bibliography{\libpath bibMarc}

\begin{thebibliography}{70}
\expandafter\ifx\csname natexlab\endcsname\relax\def\natexlab#1{#1}\fi
\expandafter\ifx\csname url\endcsname\relax
  \def\url#1{\texttt{#1}}\fi
\expandafter\ifx\csname urlprefix\endcsname\relax\def\urlprefix{URL }\fi

\bibitem[{Bather(1900)}]{bather1900.1-344}
Bather, F.~A., 1900. The {E}chinoderma. In: Lankester, E.~R. (Ed.), A treatise
  on zoology, part III. Adam \& Charles Black, London, UK, pp. 1--344.

\bibitem[{Bell(1976)}]{bell1976.1-447}
Bell, B.~M., 1976. {A study of North American Edrioasteroidea}. New York State
  Mus. Mem. 21, 1--447.

\bibitem[{Bury(1889)}]{bury1889.409-450}
Bury, H., 1889. Studies in the embryology of the echinoderms. Q. J. Microsc.
  Sci. 29~(116), 409--450.

\bibitem[{Bury(1895)}]{bury1895.45-135}
Bury, H., 1895. The metamorphosis of echinoderms. Q. J. Microsc. Sci. 38,
  45--135.

\bibitem[{Cu{\'e}not(1891)}]{cuenot1891.monografie}
Cu{\'e}not, L., 1891. {\'E}tudes morphologiques sur les {\'e}chinodermes. Arch.
  Biol. 11, 313--680.

\bibitem[{David et~al.(1994)David, Guille, Feral, and Roux}]{david1994.book}
David, B., Guille, A., Feral, J.~P., Roux, M. (Eds.), 1994. Echinoderms through
  time ({Echinoderms Dijon}). Vol.~8. A.A. Balkema, Rotterdam.

\bibitem[{David et~al.(2000)David, Lefebvre, Mooi, and
  Parsley}]{david2000.529-555}
David, B., Lefebvre, B., Mooi, R., Parsley, R., 2000. Are homalozoans
  echinoderms? {A}n answer from the extraxial-axial theory. Paleobiol. 26~(4),
  529--555.

\bibitem[{David and Mooi(1996)}]{david1996.577-584}
David, B., Mooi, R., 1996. Embryology supports a new theory of skeletal
  homologies for the phylum {E}chinodermata. C. R. Acad. Sci. Paris 319,
  577--584.

\bibitem[{David and Mooi(1998)}]{david1998.21-28}
David, B., Mooi, R., 1998. Major events in the evolution of echinoderms viewed
  by the light of embryology. In: Mooi, R., Telford, M. (Eds.), Echinoderms:
  San Francisco. Balkema, Rotterdam, pp. 21--28.

\bibitem[{David et~al.(1995)David, Mooi, and Telford}]{david1995.155-164}
David, B., Mooi, R., Telford, M., 1995. The ontogenetic basis of {L}ov{\'e}n's
  rule clarifies homologies of the echinoid peristome. In: Emson, R.~H., Smith,
  A.~B., Campbell, A.~C. (Eds.), Echinoderm research. Vol.~4. A. A. Balkema,
  Rotterdam, pp. 155--164.

\bibitem[{Delsuc et~al.(2006)Delsuc, Brinkmann, Chourrout, and
  Philippe}]{delsuc2006.965-968}
Delsuc, F., Brinkmann, H., Chourrout, D., Philippe, H., 2006. Tunicates and not
  cephalochordates are the closest living relatives of vertebrates. Nature
  439~(7079), 965--968.

\bibitem[{Gemmill(1912)}]{gemmill1912.book}
Gemmill, J.~F., 1912. The development of the starfish {{\em Solaster endeca}
  F}orbes. Trans. Zool. Soc. Lond. 20, 1--71.

\bibitem[{Gemmill(1914)}]{gemmill1914.213-294}
Gemmill, J.~F., 1914. The development and certain points in the adult structure
  of the starfish {{\em Asterias rubens}, L}. Trans. Roy. Soc. B 205, 213--294.

\bibitem[{Gemmill(1920)}]{gemmill1920.155-187}
Gemmill, J.~F., 1920. The development of the starfish {{\em Crossaster
  papposus}, M{\"u}ller and Troschel}. Q. J. Microsc. Sci. s2-64~(254),
  155--187.

\bibitem[{Giese and Pearse(1991)}]{Giese1991.book}
Giese, A.~C., Pearse, J.~S. (Eds.), 1991. Reproduction of marine invertebrates.
  Vol. VI. Echinoderms and Lophophorates. Acad. Press, New York.

\bibitem[{Gordon(1926)}]{gordon1926.259-312}
Gordon, I., 1926. The development of the calcareous test of {{\em Echinus
  miliaris}}. Phil. Trans. R. Soc. Lond. B 214, 259--312.

\bibitem[{Gordon(1929)}]{gordon1929.289-334}
Gordon, I., 1929. Skeletal development in {{\em Arbacia}, {\em Echinarachnius},
  and {\em Leptasterias}}. Phil. Trans. R. Soc. Lond. B 217, 289--334.

\bibitem[{Grass{\'e}(1948)}]{grasse1948.book}
Grass{\'e}, P.~P., 1948. Trait{\'e} de zoologie. Vol. XI. {\'E}chinodermes,
  Stomocord{\'e}s, Procord{\'e}s. Masson \& {C$^{IE}$}, Paris.

\bibitem[{Gudo(2005)}]{gudo2005.191-216}
Gudo, M., 2005. An evolutionary scenario for the origin of pentaradial
  echinoderms: implications from the hydraulic principles of form
  determination. Acta Biotheor. 53~(3), 191--216.

\bibitem[{Guensburg and Sprinkle(2001)}]{guensburg2001.131-134}
Guensburg, T.~E., Sprinkle, J., 2001. Earliest crinoids: New evidence for the
  origin of the dominant {P}aleozoic echinoderms. Geology 29~(1), 131--134.

\bibitem[{Guensburg and Sprinkle(2009)}]{guensburg2009.350-364}
Guensburg, T.~E., Sprinkle, J., 2009. Solving the mystery of crinoid ancestry:
  New fossil evidence of arm origin and development. J. Paleont. 83~(3),
  350--364.

\bibitem[{Halanych(1995)}]{halanych1995.72-76}
Halanych, K.~M., 1995. The phylogenetic position of the pterobranch
  hemichordates based on {18S} r{DNA} sequence data. Mol. Phylogeny Evol. 4,
  72--76.

\bibitem[{Hara et~al.(2006)Hara, Yamaguchi, Akasaka, Nakano, Nonaka, and
  Amemiya}]{hara2006.797-809}
Hara, Y., Yamaguchi, M., Akasaka, K., Nakano, H., Nonaka, M., Amemiya, S.,
  2006. Expression patterns of {{\em Hox}} genes in larvae of the sea lily
  {{\em Metacrinus rotundus}}. Dev. Genes Evol. 216, 797--809.

\bibitem[{Holland(1991)}]{holland1991.247-300}
Holland, N.~D., 1991. {Echinodermata: Crinoidea}. In:  \cite{Giese1991.book},
  pp. 247--300.

\bibitem[{H{\"o}rstadius(1926)}]{horstadius1926.8}
H{\"o}rstadius, S., 1926. {Embryologische Beobachtungen {\"u}ber {\em Luidia
  ciliaris} Phil., {\em L. sarsi} D{\"u}b. und Kor. und {\em Phyllophorus urna}
  G}rube. Ark. Zool. 18B~(8), 1--5.

\bibitem[{Hotchkiss(1978)}]{hotchkiss1978.537-544}
Hotchkiss, F. H.~C., 1978. Studies on echinoderm ray homologies; {L}ov{\'e}n's
  law applies to {P}aleozoic ophiuroids. J. Paleont. 52~(3), 537--544.

\bibitem[{Hotchkiss(1995)}]{hotchkiss1995.401-435}
Hotchkiss, F. H.~C., 1995. Lov{\'e}n's law and adult homologies in echinoids,
  ophiuroids, edrioasteroids, and an ophiocistioid ({Echinodermata:
  Eleutherozoa}). Proc. Biol. Soc. Washington 108, 401--435.

\bibitem[{Hotchkiss(1998)}]{hotchkiss1998.200-214}
Hotchkiss, F. H.~C., 1998. A ``rays-as-appendages'' model for the origin of
  pentamerism in echinoderms. Paleobiol. 24~(2), 200--214.

\bibitem[{Hotchkiss(2000)}]{hotchkiss2000.340-354}
Hotchkiss, F. H.~C., 2000. On the number of rays in starfish. Amer. Zool.
  40~(3), 340--354.

\bibitem[{Hyman(1955)}]{hyman1955.book}
Hyman, L.~H., 1955. The invertebrates. Vol. IV. Echinodermata. McGraw-Hill, New
  York.

\bibitem[{Jefferies(1986)}]{jefferies1986.book}
Jefferies, R. P.~S., 1986. The ancestry of the vertebrates. British Museum
  (Natural History), London.

\bibitem[{Keller et~al.(2010)Keller, Schmidt, Santella, Khairy, Bao, Wittbrodt,
  and Stelzer}]{keller2010.637-642}
Keller, P.~J., Schmidt, A.~D., Santella, A., Khairy, K., Bao, Z., Wittbrodt,
  J., Stelzer, E. H.~K., 2010. Fast, high-contrast imaging of animal
  development with scanned light sheet-based structured-illumination
  microscopy. Nat. Meth. 7~(8), 637--642.

\bibitem[{Kerr and Kim(1999)}]{kerr1999.93-103}
Kerr, A.~M., Kim, J., 1999. Bi-penta-bi-decaradial symmetry: A review of
  evolutionary and developmental trends in {H}olothuroidea ({E}chinodermata).
  J. Exp. Zool. (Mol. Dev. Evol.) 285~(2), 93--103.

\bibitem[{Komatsu et~al.(1994)Komatsu, Kawai, Nojima, and
  Oguru}]{komatsu1994.327-333}
Komatsu, M., Kawai, M., Nojima, S., Oguru, C., 1994. Development of the
  multiarmed seastar, {{\em Luidia maculata} M}{\"u}ller \& {T}roschel. In:
  \cite{david1994.book}, pp. 327--333.

\bibitem[{Lacalli(1993)}]{lacalli1993.127-133}
Lacalli, T.~C., 1993. Ciliary bands in echinoderm larvae: Evidence for
  structural homologies and a common plan. Acta Zool. 74~(2), 127--133.

\bibitem[{Lacalli and West(2000)}]{lacalli2000.421-432}
Lacalli, T.~C., West, J.~E., 2000. The auricularia-to-doliolaria transformation
  in two aspidochirote holothurians, {{\em Holothuria mexicana} and {\em
  Stichopus californicus}}. Invertebr. Biol. 119~(4), 421--432.

\bibitem[{Lartillot and Philippe(2008)}]{lartillot2008.1463-1472}
Lartillot, N., Philippe, H., 2008. Improvement of molecular phylogenetic
  inference and the phylogeny of {B}ilateria. Phil. Trans. R. Soc. B
  363~(1496), 1463--1472.

\bibitem[{Lefebvre(2003)}]{lefebvre2003.511-555}
Lefebvre, B., 2003. Functional morphology of stylophoran echinoderms.
  Palaeontology 46~(3), 511--555.

\bibitem[{Lov{\'e}n(1874)}]{loven1874.1-91}
Lov{\'e}n, S., 1874. {\'E}tudes sur les echinoid{\'e}es. Kong. Svenska Vetensk.
  Akad. Handl. 11~(7), 1--91.

\bibitem[{MacBride(1903)}]{macbride1903.285-330}
MacBride, E.~W., 1903. The development of {{\em Echinus esculentus}}, together
  with some points in the development of {{\em E. miliaris} and {\em E.
  acutus}}. Phil. Trans. B 195, 285--330.

\bibitem[{McCain and McClay(1994)}]{mccain1994.395-404}
McCain, E.~R., McClay, D.~R., 1994. The establishment of bilateral asymmetry in
  sea urchin embryos. Development 120~(2), 395--404.

\bibitem[{Mladenov and Chia(1983)}]{mladenov1983.309-323}
Mladenov, P.~V., Chia, F.~S., 1983. Development, settling behaviour,
  metamorphosis and pentacrinoid feeding and growth of the feather star {{\em
  Florometra serratissima}}. Marine Biol. 73~(3), 309--323.

\bibitem[{Mooi and David(1998)}]{mooi1998.965-974}
Mooi, R., David, B., 1998. Evolution within a bizarre phylum: homologies of the
  first echinoderms. Amer. Zool. 38, 965--974.

\bibitem[{Mooi and David(2008)}]{mooi2008.43-62}
Mooi, R., David, B., 2008. Radial symmetry, the anterior/posterior axis, and
  echinoderm {{\em Hox}} genes. Annu. Rev. Ecol. Evol. Syst. 39~(1), 43--62.

\bibitem[{Mooi et~al.(1994)Mooi, David, and Marchand}]{mooi1994.87-95}
Mooi, R., David, B., Marchand, D., 1994. Echinoderm skeletal homologies:
  classical morphology meets modern phylogenetics. In:  \cite{david1994.book},
  pp. 87--95.

\bibitem[{Mooi et~al.(2005)Mooi, David, and Wray}]{mooi2005.542-555}
Mooi, R., David, B., Wray, G.~A., 2005. Arrays in rays: terminal addition in
  echinoderms and its correlation with gene expression. Evol. Dev. 7~(6),
  542--555.

\bibitem[{Morris(2009)}]{morris2009.597-608}
Morris, V., 2009. On the sites of secondary podia formation in a juvenile
  echinoid: growth of the body types in echinoderms. Dev. Genes Evol. 219~(11),
  597--608.

\bibitem[{Morris(2007)}]{morris2007.1511-1516}
Morris, V.~B., 2007. Origins of radial symmetry identified in an echinoderm
  during adult development and the inferred axes of ancestral bilateral
  symmetry. Proc. R. Soc. B 274~(1617), 1511--1516.

\bibitem[{Morris and Byrne(2005)}]{morris2005.456-467}
Morris, V.~B., Byrne, M., 2005. Involvement of two {Hox genes and {\em Otx} in
  echinoderm body-plan morphogenesis in the sea urchin {\em Holopneustes
  purpurescens}}. J. Exp. Zool. B: Mol. Dev. Evol. 304B~(5), 456--467.

\bibitem[{Morris et~al.(2009)Morris, Selvakumaraswamy, Whan, and
  Byrne}]{morris2009.1277-1284}
Morris, V.~B., Selvakumaraswamy, P., Whan, R., Byrne, M., 2009. Development of
  the five primary podia from the coeloms of a sea star larva: homology with
  the echinoid echinoderms and other deuterostomes. Proc. R. Soc. B 276~(1660),
  1277--1284.

\bibitem[{Nakano et~al.(2003)Nakano, Hibino, Oji, Hara, and
  Amemiya}]{nakano2003.158-160}
Nakano, H., Hibino, T., Oji, T., Hara, Y., Amemiya, S., 2003. Larval stages of
  a living sea lily (stalked crinoid echinoderm). Nature 421~(6919), 158--160.

\bibitem[{Narasimhamurti(1933)}]{narasimhamurti1933.63-88}
Narasimhamurti, N., 1933. The development of {{\em Ophiocoma nigra}}. Q. J.
  Microsc. Sci. 76~(301), 63--88.

\bibitem[{Nichols(1966)}]{nichols1966.219-244}
Nichols, D., 1966. Functional morphology of the water-vascular system. In:
  Bool\-ootian, A. (Ed.), Physiology of echinodermata. Interscience, New York,
  pp. 219--244.

\bibitem[{Nichols(1972)}]{nichols1972.519-538}
Nichols, D., 1972. The water-vascular system in living and fossil echinoderms.
  Palaeontology 15~(4), 519--538.

\bibitem[{Peterson et~al.(2000)Peterson, Arenas-Mena, and
  Davidson}]{peterson2000.93-101}
Peterson, K.~J., Arenas-Mena, C., Davidson, E.~H., 2000. The {A/P} axis in
  echinoderm ontogeny and evolution: evidence from fossils and molecules. Evol.
  Dev. 2~(2), 93--101.

\bibitem[{Ritter and Crocker(1900)}]{ritter1900.247-274}
Ritter, W.~E., Crocker, G.~R., 1900. Multiplication of rays and bilateral
  symmetry in the 20-rayed starfish, {{\em Pycnopodia helianthoides}
  (S}timpson). Proc. Wash. Acad. Sci. 2, 247--274.

\bibitem[{Sauc{\`e}de et~al.(2003)Sauc{\`e}de, Mooi, and
  David}]{saucede2003.399-412}
Sauc{\`e}de, T., Mooi, R., David, B., 2003. Combining embryology and
  paleontology: origins of the anterior-posterior axis in echinoids. C. R.
  Palevol 2, 399--412.

\bibitem[{Selenka(1883)}]{selenka1883.monographie}
Selenka, E., 1883. {Die Keimbl\"atter der Echinodermen}. Entwicklungsgeschichte
  Thiere {I}~(2), 1--10.

\bibitem[{Smiley(1986)}]{smiley1986.611-631}
Smiley, S., 1986. {Metamorphosis of {\em Stichopus californicus}
  (Echinodermata: Holo\-thuroidea) and its phylogenetic implications}. Biol.
  Bull. 171~(3), 611--631.

\bibitem[{Sprinkle(1973)}]{sprinkle1973.book}
Sprinkle, J., 1973. Morphology and evolution of blastozoan echinoderms. Harvard
  University, Museum of Comparative Zoology Special Publication, Cambridge.

\bibitem[{Sprinkle and Collins(1998)}]{sprinkle1998.269-282}
Sprinkle, J., Collins, D., 1998. Revision of {{\em Echmatocrinus}} from the
  {M}iddle {C}ambrian {B}urgess {S}hale of {B}ritish {C}olumbia. Lethaia
  31~(4), 269--282.

\bibitem[{Sprinkle and Guensburg(2004)}]{sprinkle2004.266-280}
Sprinkle, J., Guensburg, T.~E., 2004. Crinozoan, blasterozoan, echinozoan,
  asterozoan, and homalozoan echinoderms. In: Webby, B.~D., et~al. (Eds.), The
  great ordovician biodiversification event. Columbia University Press, pp.
  266--280.

\bibitem[{Sumrall(1997)}]{sumrall1997.267-288}
Sumrall, C.~D., 1997. The role of fossils in the phylogenetic reconstruction of
  {E}chinodermata. Paleont. Soc. Papers 3, 267--288.

\bibitem[{Sumrall(2002)}]{sumrall2002.918-920}
Sumrall, C.~D., 2002. A new species of {{\em Anartiocystis} (Callocystitida,
  Glyptocystitida) from the Brassfield formation of central Kentucky}. J.
  Paleont. 76~(5), 918--920.

\bibitem[{Sumrall(2008)}]{sumrall2008.229-242}
Sumrall, C.~D., 2008. {The origin of Lov{\'e}n's Law in glyptocystitoid
  rhombiferans and its bearing the plate homology and the heterochronic
  evolution of the hemicosmitid peristomal border}. In: Ausich, I., Webster, G.
  (Eds.), Echinoderm Paleobiology. University of Indiana Press, Bloomington,
  pp. 229--242.

\bibitem[{Sumrall(2010)}]{sumrall2010.269-276}
Sumrall, C.~D., 2010. A model for elemental homology for the peristome and
  ambulacra in blastozoan echinoderms. In: Harris, L.~G., B{\"o}ttger, S.,
  Walker, C., Lesser, M. (Eds.), Echinoderms: Durham. Vol.~12. CRC Press /
  Balkema, Leiden (NL), pp. 269--276.

\bibitem[{Sumrall and Deline(2009)}]{sumrall2009.135-139}
Sumrall, C.~D., Deline, B., 2009. {A new species of the dual-mouthed
  paracrinoid \emph{Bistomiacystis} and a redescription of the Edrioasteroid
  \emph {Edrioaster priscus} from the upper Ordovician Curdsville member of the
  Lexington limestone}. J. Paleont. 83~(1), 135--139.

\bibitem[{Sumrall and Wray(2007)}]{sumrall2007.149-163}
Sumrall, C.~D., Wray, G.~A., 2007. Ontogeny in the fossil record:
  diversification of body plans and the evolution of ``aberrant'' symmetry in
  {P}aleozoic echinoderms. Paleobiol. 33~(1), 149--163.

\bibitem[{von Ubisch(1913)}]{ubisch1913.119-156}
von Ubisch, L., 1913. {Die Anlage und Ausbildung des Skeletsystems einiger
  Echiniden und die Symmetrieverh{\"a}ltnisse von Larve und Imago}. Zeitschr.
  Wiss. Zool. 104, 119--156.

\bibitem[{Westheide and Rieger(2007)}]{westheide2007.book}
Westheide, W., Rieger, R.~M., 2007. Spezielle {Z}oologie, Teil 1: {E}inzeller
  und {W}irbellose {T}iere, 2nd Edition. Spektrum Akademischer Verlag,
  Heidelberg.

\end{thebibliography}

\end{document}